\documentclass[debug]{epl}
\usepackage{epsfig}
\title{Elasticity-driven Nanoscale Texturing in \\  
Complex Electronic Materials
}
\shortauthor{A.R. Bishop {\it et al.}}
\shorttitle{Elasticity-driven Nanoscale Texturing}
\author{A.R. Bishop\inst{1}
T. Lookman\inst{1}, 
A. Saxena\inst{1}
\and S.R. Shenoy\inst{2} }
\institute{
\inst{1}Theoretical Division, Los Alamos National Laboratory,  
Los Alamos,NM, 87544 USA\\
\inst{2}International Centre for Theoretical Physics, Trieste 34014,
Italy \\
}

\begin{document}
\maketitle
\begin{abstract}
Finescale probes  of many complex electronic materials have
revealed a non-uniform nanoworld of sign-varying textures in strain,
charge and magnetization, forming meandering ribbons, stripe
segments or droplets. We introduce and simulate a
Ginzburg-Landau model for a  structural transition,
with strains coupling to charge
and magnetization. Charge doping acts as a local stress that
deforms surrounding unit cells without generating defects.
This seemingly innocuous constraint of elastic `compatibility',
in fact induces crucial anisotropic long-range forces of unit-cell
discrete symmetry, that interweave
opposite-sign competing strains to produce polaronic
elasto-magnetic textures in the composite variables.
Simulations with random local doping below the solid-solid
transformation temperature reveal rich multiscale texturing from induced
elastic fields: nanoscale phase
separation, mesoscale intrinsic
inhomogeneities, textural cross-coupling to external stress and
magnetic field, and temperature-dependent percolation. We describe how
this composite
textured polaron concept can be valuable for doped
manganites, cuprates and other complex electronic materials.
\end{abstract}

High resolution microscopies of many classes of
complex electronic materials such as cuprates, manganites,
ferroelastic martensites,
and relaxor
ferroelectric
titanates
\cite{Stripes,STM,Bianc,Sinha} have
revealed  previously unsuspected, and puzzling,
multiscale modulations of charge, spin,
polarization and strain variables in  stripe- or
droplet-like patterning over 1-100 nm scales, or up to hundreds of
lattice
spacings. 
 These
sign-varying inhomogeneities
or
`texturings'
fundamentally affect {\it local} electronic, magnetic and structural
properties, and
appear to be {\it intrinsic}:
arising from the coupling between degrees-of-freedom, rather than
induced by extrinsic disorder.

Ferroelastic  martensitic
alloys like FePd, and the more complex cuprates and manganites, typically
have displacive first-order structural
transitions  \cite{Stripes}, described by  {\it
symmetry-adapted} strain
tensor components as order-parameters (OP),
with  1-10 nm
criss-cross 'tweed' strain variations
above,
and 10-100 nm  `twins'
below, the structural transition \cite{Stripes}. There
is evidence for lattice/charge/spin
coupling in both cuprates and manganites in:
isotope effects \cite{Zhao};
doping-dependence of transition temperatures \cite{Stripes};
polaron signatures
with
\cite{Sinha} temperature-
and
field-dependent four-lobe signatures in diffuse X-ray and neutron
scattering; field-dependent wandering ribbons of high
conductivity
\cite{Stripes,STM,Dagotto}
; and
(fluctuating or static) stripe bubbles of
modulated buckling-angle
and spin \cite{Bianc}.
A strain-magnetization coupling
is indicated by {\it cross}-responses \cite{Stresso}: e.g.,
 manganites that show colossal
magneto-resistance (CMR) in magnetic fields
also show
colossal `stresso-resistance' (CSR) under hydrostatic pressure.

 The central questions are
conceptual:
what excitations result,  when doped local charges in a (para- or
antiferro-) magnetic
background couple to a nonlinear elastic lattice that supports a
structural transition? What is the origin of multiscale intrinsic
inhomogeneities?

We adopt a Ginzburg-Landau (GL) approach, with three generic model
assumptions:
(1) The free energy is harmonic in the non-order-parameter strain
components and  nonlinear in the order-parameter
strain (with multiple wells just above, and below, structural transition
temperatures);
(2) The charge density and magnetization variables couple locally, in
symmetry-allowed ways, to
different symmetry-adapted  strain-tensor
components, (and can then act like local internal stresses or
temperatures); (3) The charge carriers deforming a unit-cell do not damage
the
lattice by generating defects such as dislocations or vacancies
(so there is a smoothly compatible matching to the farther, and
decreasingly-strained, unit-cells).
Long-range {\it isotropic} Coulomb potentials, with or  without, or with
electron-phonon coupling, have been invoked to explain nanoscale charged
stripe patterns \cite{Zaanen}.
However, our central point here is that
charges and spins locally coupled to strains of a nonlinear lattice,
can induce multiscale mutual texturings through 
{\it anisotropic long-range}
strain-strain  forces (themselves Coulombic at a chemical-bond level
\cite{Zener}), that arise from
generic {\it elastic
compatibility} constraints \cite{Elastbook}.

{\bf GL model}: {\it 1. Strain Free Energy:}
We consider a two-dimensional (2D) first-order
square-to-rectangular transformation as a
surrogate for 3D tetragonal-orthorhombic \cite{Stripes}
structural transitions; extensions to 3D are possible
\cite{OurRC}. The
Cauchy strain tensor $\underline{\underline{E}}$ linear in
the displacement $\vec{u}$, is
$E_{\mu \nu}=(1/2)(\Delta_{\mu}u_{\nu}+\Delta_{\nu}u_{\mu})$,
where $\Delta_{\mu}$ is a discrete derivative in the $\mu = {x,y}$
directions, and small 'geometric' nonlinearities are ignored.
The symmetry-adapted strains are
the 'rectangular' or deviatoric $\varepsilon\equiv(1/\sqrt{2})(E_{xx}
-E_{yy})$, compressional $e_1=(1/\sqrt{2})(E_{xx} + E_{yy})$, and shear
$e_2= E_{xy}$, strains, respectively. The invariant free energy
is
$$F=F^{(1)}(\varepsilon,e_1,e_2) +
F^{(2)}(m)+F^{(coupling)}(n,m,\varepsilon,e_1), \eqno(1) $$
\noindent with all quantities scaled  \cite{OurRC}
to be dimensionless.
The strain contribution is \\
$F^{(1)}= (a_0/2)\sum_{\vec{r}} (\vec{\Delta} \varepsilon)^2
+F_{0}(\varepsilon)
+ F_{cs}(e_1,e_2), $
where $\sqrt{a_0}$ is a strain variation length. The
Landau term is sixth-order in the deviatoric OP strain
$F_{0}=(\tau-1)\varepsilon^ 2
+\varepsilon^2(\varepsilon^2-1)^2$. Here $\tau(T)= (T - T_{sc})/
(T_s -T_{sc})$ is a scaled temperature, and for $4/3> \tau > 0$,
$F_{0}$ has triple wells that become degenerate at $\tau(T
= T_s) =1$, reducing to  double wells for $T < T_{sc}$ or
($\tau < 0$).
The compression/shear ($cs$) terms are harmonic,
$F_{cs}=\sum_{\vec{r}}\frac{1}{2}A_1e_1^2 (\vec{r})+\frac{1}{2}A_2e_2^2
(\vec{r})$.

{\it 2. Charge and Magnetization Free Energy:}
The free energy for a magnetization variable $m(\vec{r})$  is
$F^{(2)}(m)=\sum_{\vec{r}}
f[(\frac{a_m}{2})(\vec{\Delta}m(\vec{r}))^2+(T-T_{cm})m^2(\vec{r})  +
\frac{1}{2}m^4 (\vec{r})] - h m.$
Here $f$ is a magnetic/elastic energy ratio, $h$ ($\sqrt{a_m}$) is a
magnetic field (length), and the zero-doping
magnetic transition temperature $T_{cm} < T_s$. The
symmetry-allowed couplings are
$$F^{(coupling)}(n,m,\varepsilon,e_1)=\sum_rA_{n
\varepsilon}n\varepsilon^2+
A_{nm}nm^2 + [A_{n1} n e_1 + A_{m1} m^2 e_1 +p_{1}(\vec{r}) e_1 ] ,
\eqno(2) $$
\noindent where $p_1(\vec{r}) > 0$ is an external compressional
stress. The $A_{n\varepsilon} n {\varepsilon}^2$ term
is like a local temperature  $\sim \tau \varepsilon^2$, and
for $A_{n\varepsilon} > 0$,
the charge favours
the high-temperature $\varepsilon =0$ unit-cell
symmetry. The $A_{m1} m^2 e_1$ term acts like a local
stress ($\sim m^2$) for $e_1$, and like a local
temperature ($\sim e_1$) for
$m^2$. We choose $A_{m1} > 0$, favouring  $m^2$ when  spins are closer
($e_1 < 0$).
The $m^2$ coefficient in $F$ defines an effective local
temperature deviation $\tau_{cm} (T,\vec{r}) \equiv f [T - T_{cm}^{eff}
(\vec{r})]/A_{m1} \equiv e_1 + [f(T-T_{cm}) + A_{nm} n]/A_{m1}
\equiv \delta e_1 (T,\vec{r})$.
This implies a temperature-dependent percolation: on
warming, the compressed $m \neq 0$ regions below transition (
$\tau_{cm}(T,\vec{r})= \delta e_1 (T,\vec{r}) < 0$) will shrink.
We now {\it mimic} manganites and cuprates by
further model choices
of parameter signs and interpretations of variables.

(i) For `manganites', with a zero-doping nonferromagnetic $T_{cm}=0$ parent compound,
 we focus for simplicity on the magnetization-inducing {\it mobile} electrons
produced by doping, of concentration $x_e
= <n>$ where the  number density $n(\vec{r})= \sum_i ({\kappa^2}/2\pi)e^{-\kappa
|\vec{r}-\vec{r}_i|}$ is  a sum
over normalized single-charge profiles at sites $i$,
with $2/ \kappa$ the intersite tunneling length. Here the  $A_{nm} n
m^2$ term with  $A_{nm} < 0$ means that mobile electrons lock onto and
align the ferromagnetic $m(\vec{r})$ core spins (mimicking a double
exchange/ Hund's rule, term). The mobile electrons can induce an 
effective local \cite{Dagotto} Curie temperature, $T_{cm}^{eff}
(\vec{r}) = (-A_{nm}n-A_{m1}
e_1)/f > 0$.
Since metallicity shrinks unit-cell volume,
we take the  $A_{n1} n
{e_1}$ term with $A_{n1} > 0$:  $n(\vec{r})$
favours compression,
$e_1 <0$.

(ii) For `cuprates', $n(\vec{r})$ is the local {\it
hole} number
density of doping fraction  $x_h = <n>$ into
an
antiferromagnetic parent compound with N\'eel temperature
$T_{cm} \neq 0$. With $A_{nm}
>0$ in
$A_{nm}n m^2$, the {\it staggered} magnetization
$m(\vec{r})$ is reduced by hole
doping. As the structural transition temperature also decreases rapidly
with
doping \cite{Stripes}, we take $A_{n \varepsilon} >>
1$.
Since holes repel
lattice ions, the $A_{n1} n e_1$ term has $A_{n1} < 0$:
$n(\vec{r})$ favours expansion, $e_1 > 0$.

{\it 3. Elastic Compatibility and Anisotropic Long-range Potentials}: The
St.
Venant's compatibility condition \cite{Elastbook} expresses the
no-defect constraint as  $\vec{\Delta}
\times(\vec{\Delta}\times\underline{\underline{E}})^{\dagger}=0$.
(This is analogous to a $\vec{\Delta}.\vec{B}=0$ no-monopole
condition in electromagnetism.)
In 2D we have \cite{Kartha}
$\vec{\Delta}^2e_1-\sqrt{8}\Delta_x\Delta_ye_2=
({\Delta_x^2}-{\Delta_y^2})\varepsilon$, linking OP and non-OP strains.
Minimizing the free energy $F_{cs}$ with respect to $e_{1,2}$
while maintaining the compatibility constraint, we find $e_{1,2}({\vec
k})= B_{1,2}({\vec k})\varepsilon({\vec
k})$, where $B_1 ({\vec k})= [{k_x}^2 -{k_y}^2] k^2/[k^4
+ (8 A_1/A_2) (k_x k_y)^2]$ and $B_2 (\vec{k}) = -(A_1/A_2)[\sqrt{8} k_x
k_y/k^2] B_1 (\vec{k})$.
 Substituting
back, the seemingly innocuous harmonic $F_{cs}$ yields
\cite{OurRC,Kartha}
the crucial ALR potential, encoding unit-cell fourfold symmetries,
$$F_{cs}= ({A_1}/2) \sum_{\vec{k}} U({\hat{ k}}){\vert{\varepsilon({\vec
k})}\vert} ^2; ~~~U(\hat{k}) =
[({k_x}^2 -{k_y}^2)^2/[k^4 +
({8 A_1}/ A_2) ({k_x} {k_y})^2],
 \eqno(3)$$

\noindent  while (2) yields nonlocal
couplings to the OP. Here $U(\hat{k})$
depends  on the direction $\hat{k}$  and clearly favors $\hat{k_x}
=\hat{k_y}$  diagonal strain textures, with a Meissner-like $e_{1,2}=0$
expulsion  \cite{OurRC}. In coordinate space, with $\hat{r}.\hat{r'}=
\cos (\theta - \theta')$, the potential
$U(\vec{r}-\vec{r}')\sim\cos4(\theta-\theta')/|\vec{r}-\vec{r}'|^D$
has sign-variation supporting elastic frustration, with the OP strain 
at a point receiving conflicting ("ferro/antiferro") instructions from
other surrounding strains. The power-law ($D=2$) decay arises from $U(\hat{k})$ being 
scale-free ($|\vect{k}|$-independent) at
long wavelengths,
rather than from proximity to some critical point.

Our central physical idea is quite simple. Suppose, among
(symmetry-broken)  rectangular unit  cells $\varepsilon (\vec{r}) =1$,
that a single
unit-cell is made square, $\varepsilon (\vec{r}) =0$. To maintain lattice
integrity, the neighboring (and further) unit cells must
also deform, with an admixture of non-OP strains.
As shown in simulations of a nonlinear-strain
model under local external stress, for $A_{1,2} >> 1$ the large non-OP
energy costs
can make it profitable to 
locally summon up the (degenerate) competing structure,  
in energy-lowering higher elastic multipoles: a process of  {\it adaptive elastic
screening } \cite{OurRC}.
Thus a charge, acting
as a local internal stress, can produce an unusual
sign-varying (i.e. textured) extended polaron modulated by the
anisotropic  $U(\vec{r} -\vec{r'})$, with
coupled fields like
$m(\vec{r})$ also
sign-varying. This
{\it p}olaronic {\it e}lasto-{\it m}agnetic {\it t}exture, or
`{\it pemt}on'
arises from compatibility and
competing ground states, and differs from the more familiar
magneto-elastic polaron \cite{ARBrev}
that deforms a single lattice structure.

We now choose parameters. For the martensite
$FePd$,  physical values \cite{Kartha} can be scaled \cite{OurRC} to be
dimensionless, and
$\varepsilon= e_1= e_2= 1$ correspond to strains $\sim 0.02$; the scaled
stress $p_1 =1$
(magnetic field $h = 1$) corresponds to  $
\sim 0.02$ GPa ( $\sim 0.25$
Tesla); and the non-OP and OP elastic constants are  $A_1 = 150,
A_2 = 300$. We take as illustrative,
$A_1 = 50, A_2 = 105, \sqrt{a_0} =0.5, \sqrt{a_m} = 1,{\kappa}= 2, T_s =
1,T_{sc}=0.8, f= 0.3, A_{m1} =+5$, and specific
`manganite' (`cuprate') model parameter sets as
$T_{cm} = 0, A_{nm} = -1, A_{n1} = +5,
A_{n \varepsilon}= +2$  ($T_{cm}= 0.6, A_{nm}= + 9,
A_{n1}= -5, A_{n \varepsilon} = 20$).
This is a regime of globally weak magnetism, relatively strong
electron-phonon coupling, and dominant compatibility forces.
\vskip 0.3truecm

{\bf Simulation of Textures}:
The free energy
$F=F(n,m,\varepsilon)$  minimum is found by
the overdamped limit of a general ferroelastic
dynamics \cite{OurRC}:
$$\dot{\varepsilon}=-\frac{\partial F}{\partial\varepsilon} ;~~~~
\dot{m}=-\frac{\partial F}{\partial m}.
\eqno(4) $$
\noindent
 We show  selected  relaxed
 profiles  \cite{Parent} of
$\varepsilon,e_1,m$ , with both $\vec{k}$ and $\vec{r}$
plots needed for a full understanding.
Figure 1 shows strain plots in coordinate space
(of $e_1(\vec{r})$) and in Fourier space (of $|\epsilon(\vec{k})|^2$) due
to a single charge for `manganite' parameters.
The $n e_1$ local stress term would by itself produce a
bare single-sign strain, so adaptive elastic screening is responsible for
the observed textured polaron or 'pemton'. The butterfly-like
quadrupolar (and essentially cancelling) lobes of both signs
in strain and magnetization,
extends over $\sim 20$ lattice spacings,  explicitly illustrating the
concept \cite{Stripes,Bianc,Dagotto} of  magnetic and structural \cite{STM}
`nano-scale phase separation'.
Bi-pemtons from nearby charges form stripe-like
segments.

For increasing random doping at $T= 0.5$,
the average magnetization $<m>$
rises sharply from zero  for $x_e > 0.13$, to e.g.,  $<m> = -0.21$ at
$x_e =0.15$, with symmetry-breaking in the $m(\vec{r})$ carried by the pemtons. 
Fig. 2 shows the mutually deforming 
multi-pemtons, forming
meandering ribbons of expanded/compressed
unit-cell strain  $e_1(\vec{r})$  or  oriented spins  $m
(\vec{r})$.
Fourier space plots,
e.g. of $|\varepsilon(\vec{k})|^2$ (or $|e_1(\vec{k})|^2$), show a
four-lobe shape (as in the single-pemton case) reminiscent
of diffuse X-ray and neutron scattering \cite{Sinha,Parent}.
A signature of
compatibility forces is the  $\vec{k} \rightarrow - \vec{k}$ inversion
symmetry  (in the squared strain)
relating most of  the even finescale crinkles. Fig. 2 also shows a
reduction, on warming, of both $<m>$ and of
$m(\vec{r})$ percolation\cite{STM,symmbreak}, at fixed doping.

Figure 3 shows the effects on the {\it composite} multipemton of an
external spatially varying
compressional stress $p_1(\vec{r})$ or magnetic field $h(\vec{r})$
, with four quadrants in each  picture from the long wavelength modulation. The
diagonal plots show the $p_1 \rightarrow e_1, h \rightarrow m$ or direct
responses.
The off-diagonal plots show the $p_1 \rightarrow m, h \rightarrow e_1$ or
{\it cross} responses
, e.g.  fine cloud-like gradations
reminiscent of STM/TEM pictures
\cite{STM}; and  enhanced magnetic percolation under
stress \cite{Stresso}.
The CMR/CSR
`colossal' effects can be understood  through
the ``compressed $\sim$ magnetic $\sim$ metallic"
interconnections, with fields/stresses locally tipping the delicate
balance between opposite-sign large values (see vertical scales) of
$m(\vec{r}),e_1(\vec{r})$, thus opening up  conducting channels both in
magnetization (via double exchange), or in compression
(via enhanced tunneling). The rich phase diagram \cite{Stripes} of charge-
(or pemton-) ordered states could arise from the orienting long-range
compatibility forces  \cite{OrdCh}.
Charge profile relaxation into locally compressed regions would  describe
orbital ordering, or coexisting localized/extended electronic states.


For  `cuprate' parameters, the
N\'eel temperature is  $T_{cm} = 0.6$,
and at $T= 0.5$, the parent compound has a uniform $m$.
Figure 4 shows that a `cuprate' single pemton is smaller, stronger, and
sharper than Fig.1.
The `cuprate' multi-pemton with  $x_h =
0.1$ aligns  $\varepsilon (\vec{r})$
into diagonal parallel ribbons, forming bubbles of
stripes \cite{Bianc}, with changes for higher doping.

In summary, the central insight from our model is that under
doping perturbation , a
nonlinear lattice can produce 'intrinsic inhomogeneities', that are not
quenched-defect random spots, but rather, self-organized
annealed-texture responses, induced by the multiscale effects of
local lattice-integrity constraints. These {\it composite}
textures vary with $T,p_1$ or $h$.

The compatible, {\it inter}-cell large-strain texturings must be
supported by {\it intra}-cell  deformations
of the atomic bases (`microstrain'), reflected in bond angle/length
distributions \cite{Bianc,Zener}, and so 
will be relevant for complex
electronic oxides, with atomic bases of tiltable
perovskite octahedra (that have directionally bonded transition-metal
ions, and polarizable/deformable oxygens \cite{Stripes}) .
Further electronic
structure studies would
allow  for both electronic and ionic optimizations.
Further theoretical work includes exploring  parameter
space  extensively, in 2D and 3D, in  overdamped or other \cite{OurRC}
regimes; adding  charge-hopping dynamics and charge-profile relaxations;
and strain-related microscopic modelling e.g. of
plane bucklings \cite{QP} induced by
octahedral tilts.
Further experimental work should include STM mapping of symmetry-adapted
strains to seek pemton signatures.


It is a pleasure to thank Seamus Davis, Carlos Frontera, and Venkat Pai
for useful conversations. This work was supported by the U.S. Department
of Energy.

\begin{figure}[h]
\epsfxsize=14cm
\epsffile{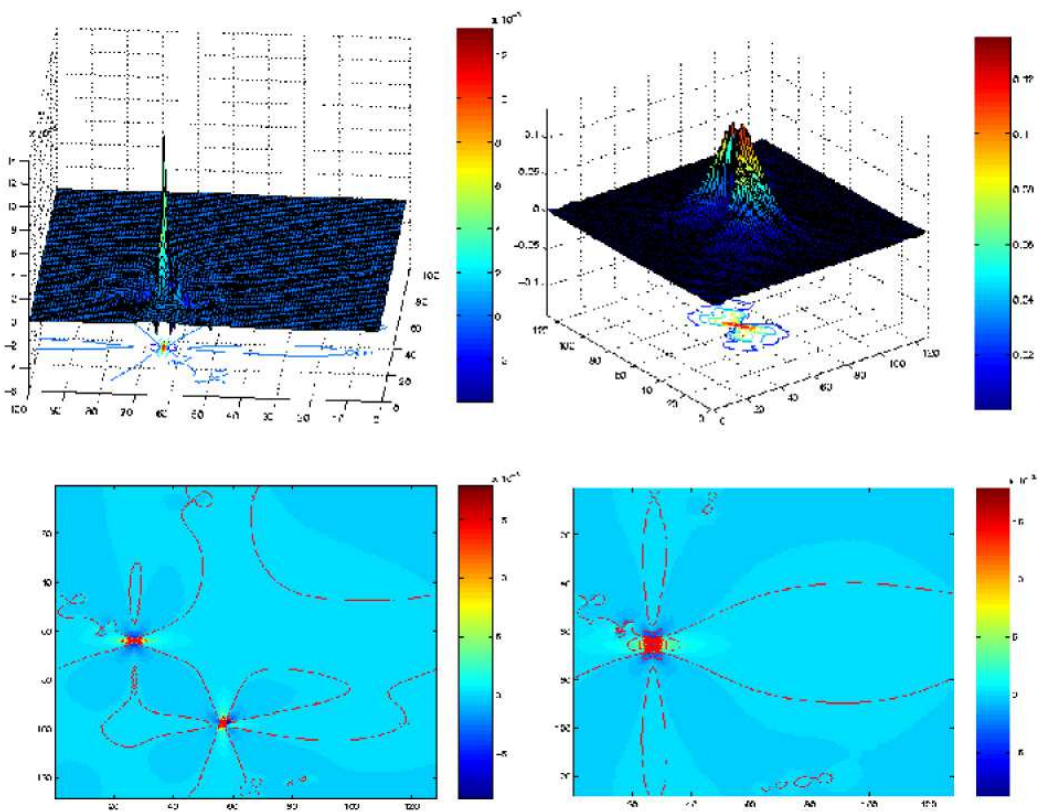}
\caption{{\it Single polaronic elasto-magnetic texture or pemton state
for `manganite' parameters:} Temperature $T= 0.5$.
All rows are read left to right. Top row: relief and contour plots of the
 compressional $e_1 (\vec{r})$
strain $\times 10^{-3}$; and  Fourier space deviatoric strain of $|\varepsilon
(\vec{k})|^2$.
Bottom row: colour plot of
$e_1(\vec{r})$ for
separated and nearby pemtons.
Note that bi-pemtons form stripe segments.}
\end{figure}

\begin{figure}[h]
\epsfxsize=14cm
\epsffile{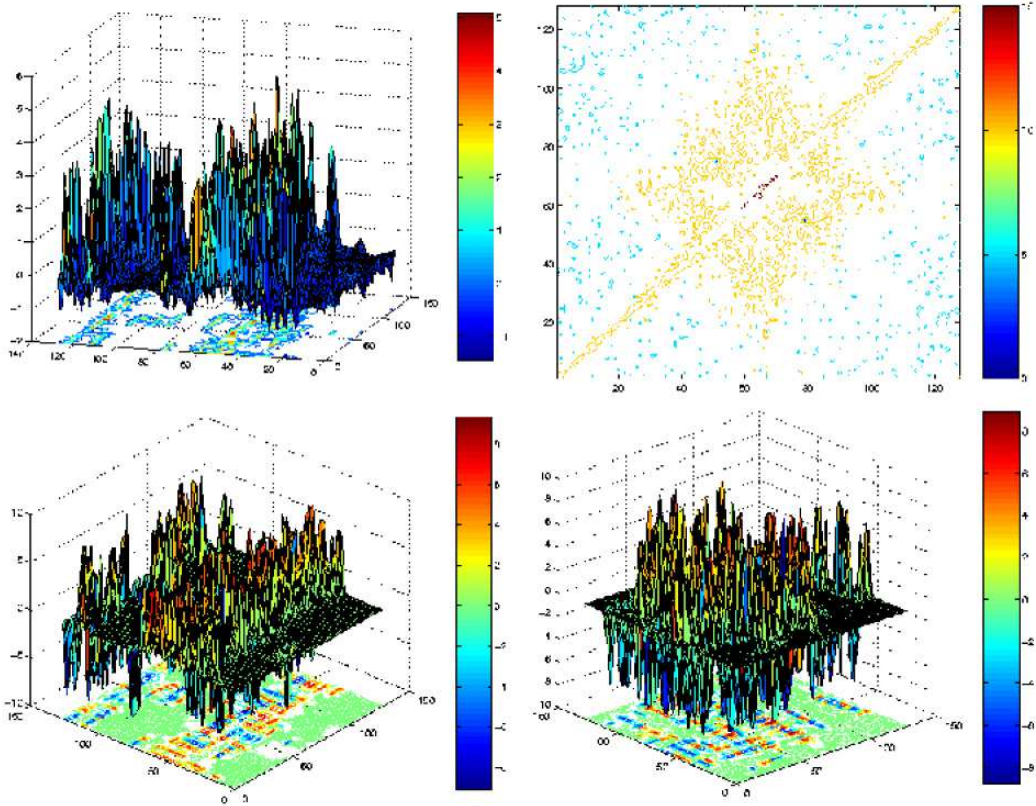}
\caption{{\it Multipemton states for `manganite' parameters:}  Electron
doping fraction  $x_e = 0.15$.
Relief/contour plots; all rows read left to right. Top row: $T= 0.5$.
Plot of the compressional
strain $e_1 (\vec{r})$.
Plot in Fourier space of {\it multiscale} $|\varepsilon (\vec{k})|^2$
(log scale).  Bottom row: magnetization
$m(\vec{r})$ for $T= 0.5$ and $0.8$, with large values compensating to
yield $<m> = -0.21$ and $-0.06$, respectively.}
\end{figure}

\begin{figure}[h]
\epsfxsize=14cm
\epsffile{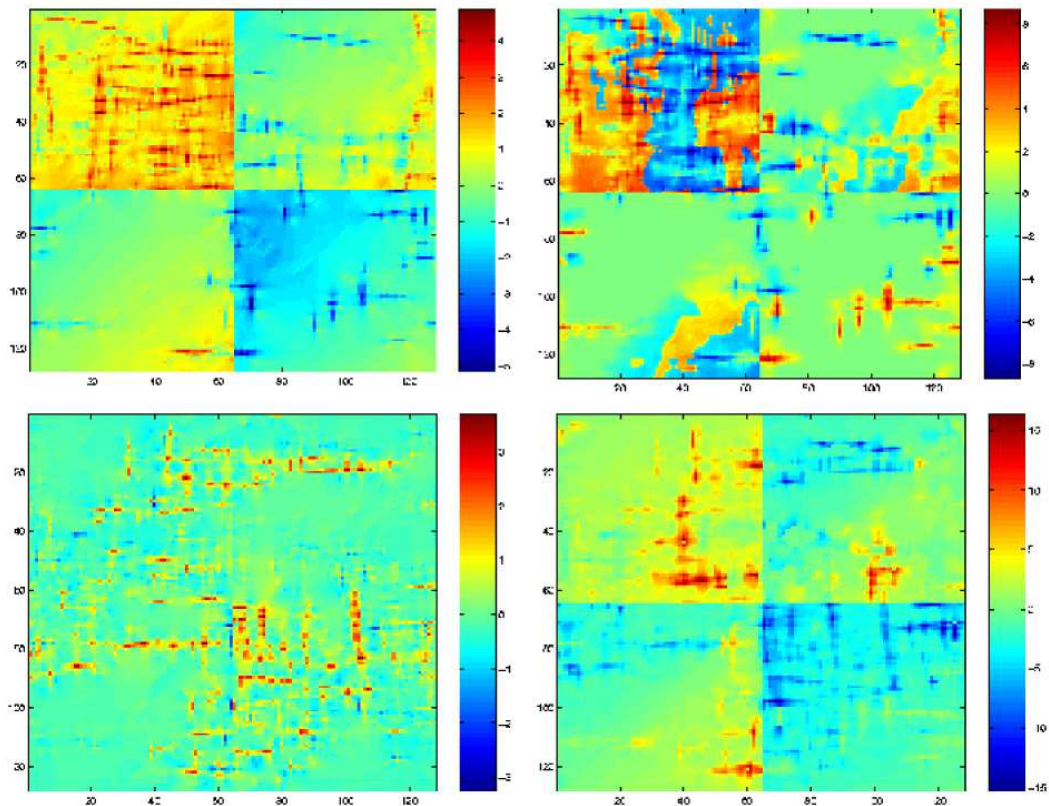}
\caption{{\it Response of `manganite' multi-pemton  states to external
stress and magnetic field:} Doping $x_e =
0.15$, and $T= 0.5$, as in the top left and bottom left of Fig. 2.
Colour plots, with all rows read left to right, with four quadrants in each, from the 
longwavelength variation of stress $p_1(\vec{r})= p_1 f(\vec{r})$ and magnetic field
$h(\vec{r})=h f(\vec{r})$, where $f(\vec{r}) = \frac{1}{2} [
\cos (2\pi x/N) + \cos (2\pi y/N)]$. Top row: $e_1 (\vec{r})$
and
$m(\vec{r})$ under stress, with
$p_1 = 20$ ($\sim 0.4 $ GPa). Bottom row: $e_1 (\vec{r})$
and
$m(\vec{r})$ in a magnetic
field, with $h = 10$ ($\sim 2.5$ tesla). Note that
$p_1(\vec{r})$ and $h(\vec{r})$ change {\it both}
$e_1$ and $m$, at {\it all} scales. }
\end{figure}

\begin{figure}[h]
\epsfxsize=15cm
\epsffile{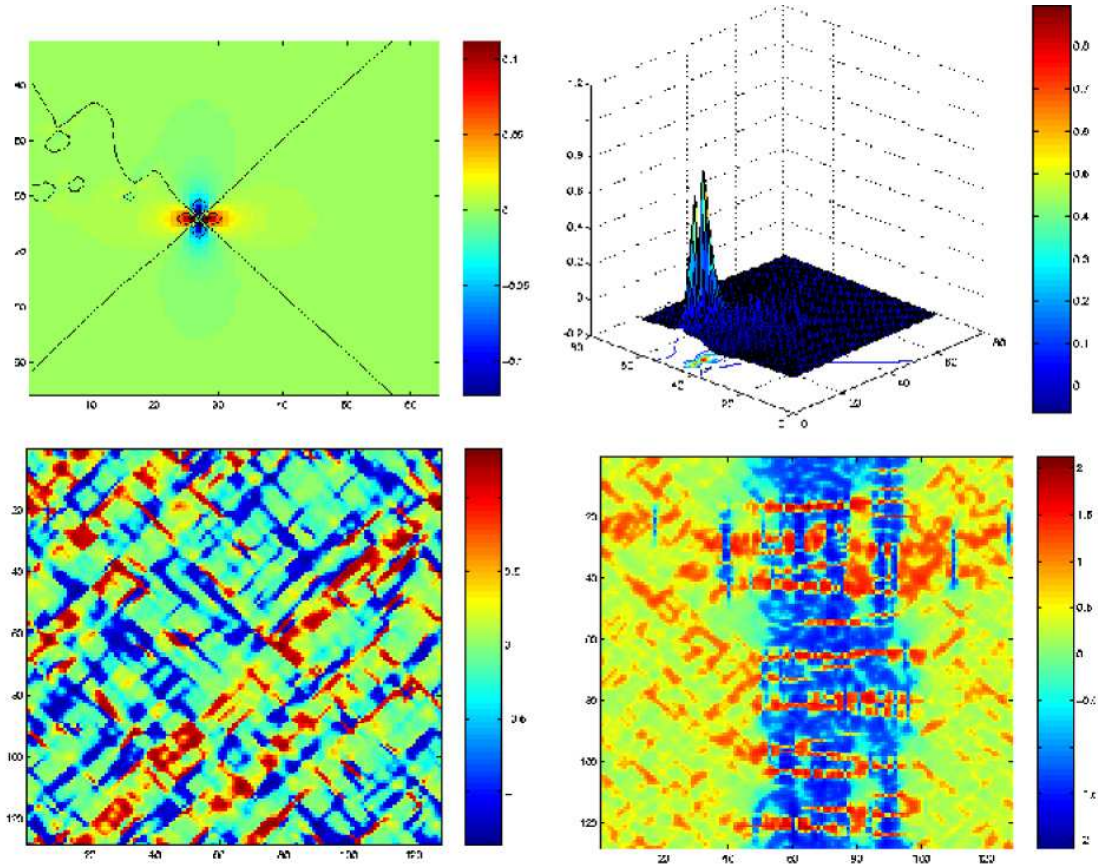}
\caption{{\it Pemtons for `cuprate' parameters:} Temperature $T=
0.5$; parent compound (subtracted) has N\'eel temperature $T_{cm} =
0.6$. All rows read left to right. Top row: Colour plot of
single-pemton $e_1 (\vec{r})$.  Relief/contour plot of staggered
magnetization $m(\vec{r})$ with  $<m> = 3 \times 10^{-3}$.
Bottom row: Colour plots of $\varepsilon (\vec{r})$ for random pemtons
with hole fractions $x_h = 0.1$ and $0.2$.}
\end{figure}


\begin{thebibliography}{99}
\bibitem{Parent}
{We consider $128 \times 128$ systems  and periodic boundary conditions,
with  charge profiles introduced through their site Fourier transforms.
We first relax the strain by itself,
from initial $\varepsilon  (\vec{r}, t=0)$
random within $\pm 0.2$. The magnetization is then coupled and relaxed
from $m(\vec{r},t=0)$ fluctuations around the zero-doping meanfield
value. This 'parent compound' has (unequal-separation) diagonal
domain walls, usually with a few metastable kinks.
Charges are added  at fixed
random positions  (but forbidding double occupancy).
Further  relaxations are up to $3 \times 10^4$ time steps, with
step size $\Delta t \sim  0.01$, using as equilibration diagnostics, the
free energy and its 'force' derivatives
, and variable averages and max/min values. The parent background is then
subtracted.
Signatures of (charge-deformed) background kinks and twins can show
up in the Figs.}
\bibitem{symmbreak}
{On cooling,
nonzero (positive and negative) $m(\vec{r})$ `metallic'
regions could possibly percolate across the
system at the same insulator-metal transition temperature. Then
symmetry-breaking to a nonzero
$<m>$ between these competing-sign backbones  could occur at a lower
ferromagnetic temperature. Alternatively, for other parameters, isolated islands could order
($<m>\neq 0$)  before  percolation, inverting the sequence of the
transitions.}
\bibitem{OrdCh}
{In our model, for periodic charge arrays, momentum conservation on
the $n,m,e_1$ couplings of (3) imply that dominant (single) wavevectors
are related by  $k_n=2k_m = k_1$.}
\end{thebibliography}
\end{document}